\documentclass[prd,superscriptaddress,amsfonts,amssymb,amsmath,showpacs,twocolumn]{revtex4-2}
\usepackage{bm}
\usepackage{amsfonts}
\usepackage{latexsym}
\usepackage{graphicx}
\usepackage{amsmath}
\usepackage{palatino}
\usepackage{mathpazo}
\usepackage{textcomp}
\usepackage{float}
\usepackage{booktabs}
\usepackage{dcolumn}
\usepackage{booktabs}
\usepackage{multirow}
\usepackage{hyperref}
\hypersetup{colorlinks,citecolor=blue}
\usepackage{amsmath}
\usepackage{xcolor}
\usepackage{orcidlink}
\usepackage{epsfig}
\usepackage{caption}
\usepackage{subcaption}
\usepackage{commath}
\def\jnl@style{\it}
\def\aaref@jnl#1{{\jnl@style#1}}

\def\aaref@jnl#1{{\jnl@style#1}}

\def\aj{\aaref@jnl{AJ}}                   
\def\apj{\aaref@jnl{ApJ}}                 
\def\apjl{\aaref@jnl{ApJ}}                
\def\apjs{\aaref@jnl{ApJS}}               
\def\apss{\aaref@jnl{Ap\&SS}}             
\def\aap{\aaref@jnl{A\&A}}                
\def\aapr{\aaref@jnl{A\&A~Rev.}}          
\def\aaps{\aaref@jnl{A\&AS}}              
\def\mnras{\aaref@jnl{Mon.~Not.~Roy.~Astron.~Soc.}}             
\def\prd{\aaref@jnl{Phys.~Rev.~D}}        
\def\prc{\aaref@jnl{Phys.~Rev.~C}}  
\def\prl{\aaref@jnl{Phys.~Rev.~Lett.}}    
\def\qjras{\aaref@jnl{QJRAS}}             
\def\skytel{\aaref@jnl{S\&T}}             
\def\ssr{\aaref@jnl{Space~Sci.~Rev.}}     
\def\zap{\aaref@jnl{ZAp}}                 
\def\nat{\aaref@jnl{Nature}}              
\def\aplett{\aaref@jnl{Astrophys.~Lett.}} 
\def\apspr{\aaref@jnl{Astrophys.~Space~Phys.~Res.}} 
\def\physrep{\aaref@jnl{Phys.~Rep.}}      
\def\physscr{\aaref@jnl{Phys.~Scr}}       
\def\commat{\aaref@jnl{Comm.~Math.~Phys.}}              
\def\science{\aaref@jnl{Science}}               
\def\cqg{\aaref@jnl{Classical Quant.~Grav.}}            
\def\jpcs{\aaref@jnl{JPCS}}                                     
\def\ijmpd{\aaref@jnl{Int.~J.~Mod.~Phys.~D}}                    
\def\grg{\aaref@jnl{Gen.~Relat.~Gravit.}}               
\def\rpp{\aaref@jnl{Rep.~Prog.~Phys.}}          
\def\npa{\aaref@jnl{Nucl.~Phys.~A}}        
\def\lrr{\aaref@jnl{Living Rev.~Rel.}}                   
\def\jcap{\aaref@jnl{J.~Cosmology Astropart.~Phys.}}    
\def\rmp{\aaref@jnl{Rev.~Mod.~Phys.}}   
\def\epjc{\aaref@jnl{Eur.~Phys.~J.~C}}

\allowdisplaybreaks[1]

\begin{document}

\color{black}       

\title{Testing the viability of $f(T, \mathcal{T})$ gravity models via effective equation of state constraints}

\author{M. Koussour\orcidlink{0000-0002-4188-0572}}
\email[Email: ]{pr.mouhssine@gmail.com}
\affiliation{Department of Physics, University of Hassan II Casablanca, Morocco.}

\author{O. Donmez\orcidlink{0000-0001-9017-2452}}
\email[Email: ]{orhan.donmez@aum.edu.kw}
\affiliation{College of Engineering and Technology, American University of the Middle East, Egaila 54200, Kuwait.}

\author{S. Bekov\orcidlink{0000-0002-8846-0604}}
\email[Email: ]{ss.bekov@gmail.com \textcolor{black}{(corresponding author)}}
\affiliation{Department of General and Theoretical Physics, L.N. Gumilyov Eurasian National University, Astana 010008, Kazakhstan.}
\affiliation{Kozybayev University,  Petropavlovsk, 150000, Kazakhstan.}

\author{A. Syzdykova}
\email[Email: ]{Syzdykova_am@mail.ru}
\affiliation{Department of General and Theoretical Physics, L.N. Gumilyov Eurasian National University, Astana 010008, Kazakhstan.}

\author{S. Muminov\orcidlink{0000-0003-2471-4836}}
\email[Email: ]{sokhibjan.muminov@gmail.com}
\affiliation{Mamun University, Bolkhovuz Street 2, Khiva 220900, Uzbekistan.}

\author{A.I. Ashirova\orcidlink{0000-0002-3399-7527}}
\email[Email: ]{ashirova_anorgul@mamunedu.uz}
\affiliation{Mamun University, Bolkhovuz Street 2, Khiva 220900, Uzbekistan.}


\begin{abstract}
This paper rigorously examines the potential of the $f(T, \mathcal{T})$ theory as a promising framework for understanding the dark sector of the universe, particularly in relation to cosmic acceleration. The $f(T, \mathcal{T})$ theory extends gravitational dynamics by incorporating both the torsion scalar $T$ and the trace of the energy-momentum tensor $\mathcal{T}$. Further, we explore the functional form $f(T, \mathcal{T}) = T + \beta \mathcal{T}$, where $\beta$ is a free parameter that modulates the matter's influence on spacetime evolution. To evaluate this model, we employ an effective EoS parameter dependent on redshift $z$, to solve the field equations and analyze the evolution of the Hubble parameter $H(z)$. Using a joint dataset ($H(z)+Pantheon^+$) and the Markov Chain Monte Carlo (MCMC) method with Bayesian analysis, we obtain the best-fit parameter values: $H_0 = 68.04 \pm 0.64$, $\beta = 0.14 \pm 0.17$, and $\gamma = 0.96^{+0.38}_{-0.69}$, which align well with current observational data. Our findings indicate a deceleration parameter of $q_0 = -0.51$, supporting a present-day accelerated expansion phase, with a transition redshift $z_t = 0.57$ marking the universe’s shift from deceleration to acceleration. Moreover, we confirm a positive cosmic fluid energy density, reinforcing stability, and find an EoS parameter value of $\omega_0 = -0.76$, consistent with quintessence-driven acceleration. These results underscore the viability of $f(T, \mathcal{T})$ as a robust framework for addressing the accelerating universe and dark energy dynamics, paving the way for future investigations into its cosmological implications.

\textbf{Keywords:} $f(T, \mathcal{T})$ gravity, cosmic acceleration, dark energy, equation of state, MCMC, quintessence, and observational cosmology.
\end{abstract}

\maketitle

\section{Introduction}\label{sec1}

Beginning with the Big Bang and the formation of fundamental elements, our understanding of the universe has evolved to include enigmatic components like dark energy (DE) and dark matter (DM), which continue to challenge modern physics. Yet, numerous theories have emerged to address these mysteries, offering potential explanations for the accelerating expansion of the universe. Coinciding with these developments, groundbreaking observations in cosmology indicate that the universe’s expansion rate is increasing, prompting deeper investigations into alternative gravitational models and the dark sector that drives this acceleration. This acceleration has been substantiated by observations from type Ia supernovae (SNe Ia) studies \cite{R1, R2}, baryon acoustic oscillations (BAO) analyses \cite{R3, R4}, findings from the wilkinson microwave anisotropy probe (WMAP) \cite{R5}, and investigations of cosmic microwave background (CMB) radiation \cite{R6, R7}. Together, these studies highlight one of the most compelling questions in cosmology today. Moreover, these findings have uncovered that a mysterious component, DE, constitutes approximately 70\% of the universe. DE is typically described by an equation of state (EoS) parameter $\omega= \frac{p}{\rho}$, representing the ratio of the spatially homogeneous pressure $p$ to the energy density $\rho$. Recent cosmological observations indicate significant uncertainties, making it challenging to distinguish between the scenarios $\omega< -1$, $\omega= -1$, and $\omega> -1$. The WMAP9 results \cite{O1}, which combine data from $H_0$ measurements, SNe Ia, CMB, and BAO, suggest $\omega_0 = -1.084 \pm 0.063$. In 2015, the Planck collaboration reported $\omega_0 = -1.006 \pm 0.0451$ \cite{O2}, and later, in 2018, they refined this value to $\omega_0 = -1.028 \pm 0.032$ \cite{O3}.

On the other hand, the geometrical theory of general relativity (GR) has provided a foundational explanation of observational evidence, leading to a significant understanding of the nature of space and time. Despite its remarkable success, recent observations have raised questions about the validity of standard GR, which may have certain limitations, especially on cosmic scales and beyond the solar system. While the cosmological constant stands as the primary candidate for DE, it is not without issues, including the coincidence problem and the fine-tuning problem \cite{R8,R9}. The coincidence problem questions why DE and matter densities are of the same order of magnitude today, despite evolving differently over time. The fine-tuning problem refers to the need for an extremely precise value of the cosmological constant to match observations, which seems unlikely given the natural scales of the universe, suggesting potential gaps in our current understanding of DE. The accelerated expansion of the universe can be approached in two ways. The first approach involves modifying the energy-momentum tensor in Einstein’s field equations through additions such as scalar fields (e.g., quintessence, phantom models) \cite{R10,R11,R12}, exotic EoS like the Chaplygin gas \cite{R13,R14}, and incorporating effects such as bulk viscosity \cite{R15,R16,R17,R18,R19}. The second approach focuses on altering the geometric framework of spacetime in Einstein’s equations, leading to what are commonly known as modified gravity theories (MGTs) \cite{R20}.

MGT is a significant field of modern cosmology that aims to develop a comprehensive framework explaining the universe's early evolution and observed late-time accelerated expansion. This field explores alternatives to GR by modifying the gravitational action, allowing for additional terms that can account for phenomena such as DE and DM without invoking new forms of matter. One of the simplest and most extensively studied approaches within this field is the $f(R)$ gravity \cite{R21,R22,R23}, which generalizes GR by replacing the Ricci scalar $R$ in the action with an arbitrary function $f(R)$. This modification introduces extra degrees of freedom that can describe a wide range of cosmological dynamics. However, the metric formulation of $f(R)$ gravity often predicts deviations from GR that conflict with observations at solar system scales, highlighting the challenge of reconciling MGT with both local and cosmological observations. To address these issues, other MGTs have been developed that incorporate non-minimal couplings between matter and curvature or other geometrical invariants. For example, $f(R, \mathcal{T})$ gravity \cite{RR24,RR25,RR26,RR27} includes an explicit dependence on the trace of the energy-momentum tensor $\mathcal{T}$, providing a mechanism for the interaction between geometry and matter. This approach aims to introduce effective gravitational behavior at cosmological scales while maintaining consistency with local tests of GR. Similarly, $f(R, G)$ gravity \cite{RR28,RR29} extends the gravitational action by including a function of the Gauss-Bonnet invariant $G$, offering another pathway to accommodate late-time cosmic acceleration without a cosmological constant. These theories represent a diverse toolkit within MGT, each seeking to balance theoretical appeal with observational compatibility in addressing fundamental cosmological questions.

Since GR is inherently a geometric theory based on Riemannian geometry, one promising direction for generalizing gravity theories involves exploring more complex geometrical structures that might capture the behavior of gravitational fields at all scales. While Riemannian geometry works well at the solar system scale, it may not fully describe gravitational interactions at larger, cosmological scales. 

This idea led to the concept of Weitzenb\"{o}ck spaces, introduced by Weitzenb\"{o}ck himself, as a broader framework for geometry \cite{RR30}. A Weitzenb\"{o}ck manifold is defined by specific properties: the metric tensor $g_{\sigma\lambda}$ is covariantly constant ($\nabla_\mu g_{\sigma\lambda} = 0$), the torsion tensor $T^\mu_{\sigma\lambda}$ is non-zero, and the Riemann curvature tensor $R^\mu_{\nu\sigma\lambda}$ vanishes. These unique properties imply that, in the absence of torsion ($ T^\mu_{\sigma\lambda} = 0$), a Weitzenb\"{o}ck manifold reduces to a Euclidean space. However, when torsion is present, it varies across different regions within the manifold, allowing for alternative ways to represent gravitational effects. One important aspect of Weitzenb\"{o}ck geometry is that its zero-curvature nature results in distant parallelism, also called absolute parallelism or teleparallelism. This characteristic caught the attention of Einstein, who applied Weitzenb\"{o}ck-type spacetimes to develop a unified theory of electromagnetism and gravity, giving rise to the concept of teleparallel gravity as an alternative to traditional curvature-based descriptions of spacetime \cite{RR31}. In the teleparallel gravity, the fundamental idea is to replace the spacetime metric $g_{\mu\nu}$, which traditionally describes gravitational properties, with a set of tetrad vectors $e_i^{\mu}$. These tetrad fields generate torsion, which can then be used to fully describe gravitational effects, effectively substituting curvature with torsion. This leads to the Teleparallel Equivalent of GR (TEGR), initially proposed in works such as \cite{RR32,RR33,RR34} and now commonly referred to as $f(T)$ gravity theory. In $f(T)$ theories, torsion completely compensates for curvature, resulting in a flat spacetime. A key advantage of $f(T)$ gravity is that its field equations are second-order, unlike $f(R)$ gravity, which in the metric formulation involves fourth-order equations. Teleparallel theories, including $f(T)$ gravity, have been extensively applied to astrophysical phenomena and cosmology, particularly in explaining the universe’s late-time accelerated expansion without invoking DE \cite{RR35,RR36,RR37,RR38,RR39,RR40,RR41,Re1,Re2,Re3,Re4,Re5,Re6,Re7,Re8}.

Recently, Harko et al. \cite{Harko_fTT} introduced an extension to teleparallel gravity by introducing the energy-momentum scalar $\mathcal{T}$ alongside the torsion scalar $T$, creating what is now commonly referred to as $f(T, \mathcal{T})$ gravity. This framework, which incorporates both the torsion scalar $T$ and the energy-momentum scalar $\mathcal{T}$, has been the subject of numerous studies that have deepened our understanding of its implications. Junior et al. \cite{TT17} investigated the reconstruction, thermodynamics, and stability of the $\Lambda$CDM model within this theory, demonstrating its compatibility with observational data and its relevance to classical and quantum gravity. Further, Harko et al. \cite{HARK} introduced a non-minimal torsion-matter coupling in $f(T)$ gravity. Moreover, Momeni and Myrzakulov \cite{TT18} conducted a study on the cosmological reconstruction of $f(T, \mathcal{T})$ gravity, revealing fundamental dynamics of the universe within this framework. Farrugia and Said \cite{TT19} analyzed the growth factor in $f(T, \mathcal{T})$ gravity, offering detailed clarifications into structure formation, while Pace and Said \cite{TT20} took a perturbative approach to explore neutron stars in this context, advancing our understanding of compact objects. In this study, we reconstruct an effective EoS parameter to investigate late-time acceleration within the framework of $f(T, \mathcal{T})$ gravity. Specifically, we consider the linear functional form $f(T, \mathcal{T}) = T + \beta \mathcal{T}$, where $\beta$ is a free parameter, and we employ an effective EoS parameter that varies with redshift $z$ as $\omega(z) = -\frac{3}{\gamma (z+1)^3 + 3}$ \cite{Ankan/2015}. This form allows us to solve the field equations and study the evolution of the Hubble parameter $H(z)$. We constrain the model parameters using a joint dataset ($H(z) + Pantheon^+$), applying the Markov Chain Monte Carlo (MCMC) method along with Bayesian analysis.

This paper is organized as follows. Sec. \ref{sec2} covers the basics of $f(T, \mathcal{T})$ gravity. In Sec. \ref{sec3}, we discuss the formulation of the EoS parameter within $f(T, \mathcal{T})$ gravity and derive an expression for the Hubble parameter. The observational constraints and data analysis to constrain the model parameters are presented in Sec. \ref{sec4}. Sec. \ref{sec5} includes a discussion of the behavior of cosmological parameters, such as the deceleration parameter, energy density, and the EoS parameter. Finally, Sec. \ref{sec6} provides conclusions and future prospects.

\section{Basics of $f(T, \mathcal{T})$ gravity}\label{sec2}

In this paper, Greek indices represent the spacetime coordinates, while Latin indices correspond to the tangent space. The primary field, known as the vierbein or tetrad, $\mathbf{e}_A(x^\mu)$, is a set of four linearly independent vector fields that provide a local orthonormal basis for the tangent space at each point $x^\mu$ in spacetime. It satisfies the relation $\mathbf{e}_A \cdot \mathbf{e}_B = \eta_{AB}$, where $\eta_{AB} = \text{diag}(1, -1, -1, -1)$ is the Minkowski metric. The vierbein relates the local tangent space to the spacetime coordinates, allowing us to express it in terms of its components as $\mathbf{e}_A = e^\mu_A \partial_\mu$. Hence, the metric tensor can be written as
\begin{equation}
\label{metrdef}
g_{\mu\nu}(x)=\eta_{AB}\, e^A_\mu (x)\, e^B_\nu (x).
\end{equation}

In the teleparallel approach, the components of the vierbein are parallelized across different points in spacetime, meaning they remain consistent in orientation relative to each other throughout the manifold. This property gives rise to the term "teleparallel", which implies "distant parallelism" or a structure without traditional curvature. This framework employs the Weitzenb\"{o}ck connection, defined as $\Gamma^\lambda_{\nu\mu}\equiv e^\lambda_A\:
\partial_\mu
e^A_\nu$ \cite{RR30}. The Weitzenb\"{o}ck connection differs from the Levi-Civita connection in that it is constructed purely from the vierbein and its derivatives, resulting in zero curvature but a non-zero torsion. This torsion captures gravitational effects in teleparallel gravity, as opposed to curvature in GR,
\begin{equation}
\label{torsion2}
{T}^\lambda_{\:\mu\nu}=\Gamma^\lambda_{
\nu\mu}-%
\Gamma^\lambda_{\mu\nu}
=e^\lambda_A\:(\partial_\mu
e^A_\nu-\partial_\nu e^A_\mu).
\end{equation}

Furthermore, we introduce the contorsion tensor, defined as
\begin{equation}
K^{\mu\nu}{}_{\rho}\equiv-\frac{1}{2}\Big(T^{\mu\nu}{}_{\rho}
-T^{\nu\mu}{}_{\rho}-T_{\rho}{}^{\mu\nu}\Big),    
\end{equation}
where $T^{\mu\nu}{}_{\rho}$ represents the torsion tensor. The contorsion tensor quantifies the degree of torsion present in spacetime and is instrumental in relating the geometric properties of teleparallel gravity to the conventional framework of GR. Further, we define the tensor
\begin{equation}
S_{\rho}{}^{\mu\nu}\equiv\frac{1}{2}\Big(K^{\mu\nu}{}_{\rho}
+\delta^\mu_\rho
\:T^{\alpha\nu}{}_{\alpha}-\delta^\nu_\rho\:
T^{\alpha\mu}{}_{\alpha}\Big).    
\end{equation}

From the above equations, the torsion scalar can be obtained \cite{Arcos/2004,Maluf/2013},
\begin{equation}
\label{torsionscalar}
T\equiv\frac{1}{4}
T^{\rho \mu \nu}
T_{\rho \mu \nu}
+\frac{1}{2}T^{\rho \mu \nu }T_{\nu \mu\rho}
-T_{\rho \mu}{}^{\rho }T^{\nu\mu}{}_{\nu}.
\end{equation}

Therefore, by incorporating the torsion scalar $T$ into the action and varying it with respect to the vierbeins, the resulting equations are equivalent to those of GR. This approach establishes the framework known as the TEGR, which serves as a foundational basis for developing various modifications of gravitational theory. A significant modification is the addition of $T + f(T)$ in the action, which leads to the formulation of $f(T)$ gravity. Further, the action for $f(T, \mathcal{T})$ gravity, which extends the $f(T)$ theory, is given by \cite{Harko_fTT}
\begin{equation}
S= \frac{1}{16\,\pi\,G}\,\int d^{4}x\,e\,\left[T+f(T,\mathcal{T})\right]%
+\int d^{4}x\,e\,\mathcal{L}_{m}.
\label{action1}
\end{equation}%

Here, $e = \det(e^A_\mu) = \sqrt{-g}$, where $G$ represents Newton's constant, and $f(T, \mathcal{T})$ is an arbitrary function of the torsion scalar $T$ and the trace $\mathcal{T}$ of the energy-momentum tensor $\mathcal{T}_{\rho}{}^{\nu}$. In addition, $\mathcal{L}_{m}$ denotes the matter Lagrangian density. From this point forward, we adopt the standard convention that $\mathcal{L}_{m}$ depends only on the vierbein and not on its derivatives. By varying the action (\ref{action1}) with respect to the vierbeins, the resulting field equations are given by
\begin{multline}
\left(1+f_{T}\right) \left[e^{-1} \partial_{\mu}{(e
e^{\alpha}_{A}
S_{\alpha}^{~\rho \mu})}-e^{\alpha}_{A} T^{\mu}_{~\nu \alpha} S_{\mu}^{~\nu
\rho}\right]+\\ 
\left(f_{TT} \partial_{\mu}{T}+f_{T\mathcal{T}} \partial_{\mu}{%
\mathcal{T}}\right) e^{\alpha}_{A} S_{\alpha}^{~\rho \mu}+ e_{A}^{\rho}
\left(\frac{f+T}{4}\right)-\\
f_{\mathcal{T}} \left(\frac{e^{\alpha}_{A} \mathcal{T}%
{}_{\alpha}^{~~\rho}+p e^{\rho}_{A}}{2}\right)=4\pi G e^{\alpha}_{A}
\mathcal{T}_{\alpha}{}^{\rho},
\label{geneoms}
\end{multline}
where $f_{\mathcal{T}}=\partial{f}/\partial{\mathcal{T}}$ and
$f_{T\mathcal{T%
}}=\partial^2{f}/\partial{T} \partial{\mathcal{T}}$.

Let's now consider the Universe described by the homogeneous, isotropic, and spatially flat Friedmann-Lemaître-Robertson-Walker (FLRW) metric. Here, homogeneous means that the properties of the Universe are the same at every point in space, while isotropic implies that the Universe looks the same in all directions. The metric can be expressed as
\begin{equation}
ds^2 = dt^2 - a(t)^2 \left( dx^2 + dy^2 + dz^2 \right),
\end{equation}
where $a(t)$ denotes the scale factor of the Universe, which quantifies the relative expansion of the Universe over time. Also, the tetrad field and torsion scalar for this metric are calculated and obtained as $e_{\mu}^A={\rm
diag}(1,a(t),a(t),a(t))$ and $T=-6 H^2$, respectively. In this context, $H$ is the Hubble parameter, indicating the rate of the Universe's expansion, defined as $H \equiv \frac{\dot{a}}{a}$.

Furthermore, we assume that the universe consists of a perfect fluid, which is characterized by an energy-momentum tensor with no shear stresses and isotropic pressure. The perfect fluid represents a combination of pressureless DM and DE contributions arising naturally from the $f(T,\mathcal{T})$ gravity framework. The energy-momentum tensor for a perfect fluid is expressed as
\begin{equation}
T_{\mu\nu} = (\rho + p)u_\mu u_\nu - p g_{\mu\nu},
\end{equation}
where $\rho$ is the energy density, $p$ is the pressure, and $u_\mu$ is the four-velocity of the fluid.

By applying the metric and the field equation (\ref{geneoms}), we derive the generalized Friedmann equations as
\cite{Harko_fTT}:
\begin{equation}\label{F1}
H^2 =\frac{8\pi G}{3}\rho - \frac{1}{6}\left(f+12H^2f_T \right)+f_\mathcal{T}\left(\frac{\rho+p}{3} \right),
\end{equation}

\begin{multline} \label{F2}
\dot{H}= -4\pi G(\rho+p)-\dot{H}(f_T-12H^2 f_{TT})\\-H(\dot{\rho}-3\dot{p}) f_{T \mathcal{T }} - f_\mathcal{T}\left(\frac{\rho+p}{2} \right).
\end{multline}

Here, the dot notation (·) indicates a derivative with respect to time and $\mathcal{T}=\rho-3 p$. 

\section{Formulation of the EoS parameter in $f(T,\mathcal{T})$ gravity}\label{sec3}

The motivation for exploring modified gravity theories, particularly the linear form, stems from the compelling need to address cosmic acceleration while remaining consistent with the successes of GR. Extensive analyses of cosmological observations, along with stringent tests conducted within the solar system, have consistently validated GR in regimes where gravitational dynamics are well understood. Consequently, any modifications to the standard framework must be subtle and rigorously constrained by empirical data. Modified gravity theories, such as $f(T, \mathcal{T})$ gravity, offer a promising avenue for extending GR, provided they adhere closely to its predictions to withstand observational scrutiny. This leads us to expect that the $f(T)$ function should take a linear form, enabling it to recover GR in appropriate limits while accommodating minor adjustments that could explain phenomena such as cosmic acceleration. For these reasons, we consider the linear functional form as a viable framework for investigating these modifications, thereby facilitating a deeper understanding of the underlying dynamics of the universe \cite{fTT1,fTT2,fTT3},
\begin{equation}\label{3e}
f(T,\mathcal{T})= T + \beta \mathcal{T}.
\end{equation}
where $\beta$ is a constant. Thus, we obtain $f_{T}= 1$, $f_{\mathcal{T}}= \beta$, $f_{TT}= 0$, and $f_{T \mathcal{T}}= 0$.

Using Eqs. (\ref{F1}) and (\ref{F2}), we obtain the expressions for pressure and energy density as
\begin{equation}
\rho=-\frac{6 \left((2 \beta +3) H^2+5 \beta  \dot{H}\right)}{4 \beta  (\beta +1)-3},
\label{rho}
\end{equation}
and
\begin{equation}
p=\frac{6 \left((2 \beta +3) H^2+(\beta +2) \dot{H}\right)}{4 \beta  (\beta +1)-3}.
\label{p}
\end{equation}

The effective (or total) EoS parameter, denoted as $\omega$, defines the relationship between pressure $p$ and energy density $\rho$ in cosmological models, expressed as $\omega = \frac{p}{\rho}$. From Eqs. (\ref{rho}) and (\ref{p}), we get
\begin{equation}
\label{w}
\omega=-\frac{(2 \beta +3) H^2+(\beta +2) \dot{H}}{(2 \beta +3) H^2+5 \beta  \dot{H}}.
\end{equation}

In addition, by using the relation $\frac{a_{0}}{a} = 1 + z$, where $z$ denotes the redshift—a measure of how much the wavelength of light has been stretched due to the expansion of the universe—we can establish the relationship between time $t$ and redshift $z$ as described below,
\begin{equation}
\label{zt}
\frac{d}{dt}= \frac{dz}{dt}\frac{d}{dz}= -(1+z) H(z)\frac{d}{dz},
\end{equation}

Here, we normalize the current value of the scale factor to $a_{0} = 1$. The Hubble parameter can then be expressed in the following form: $\dot{H}= -(1+z) H(z)\frac{dH}{dz}$.

To solve Eqs. (\ref{rho}) and (\ref{p}) for $H$ in terms of $z$, we introduce an effective EoS expressed in terms of redshift as $\omega(z) = -\frac{3}{\gamma (z+1)^3 + 3}$, where $\gamma$ is a constant \cite{Ankan/2015}. This formulation serves a dual purpose: it captures the late-time accelerated expansion of the Universe while also mimicking DM-like behavior during its earlier stages. Specifically, as $z \to \infty$, which corresponds to the early Universe when matter was dominant, $\omega$ approaches 0. This value indicates a matter-dominated era where the pressure is negligible compared to the energy density, consistent with the behavior of non-relativistic matter. At $z = 0$, the current redshift, the effective EoS takes the form $\omega(0) = -\frac{3}{\gamma + 3}$. This value provides a measure of the present-day dynamics, which is essential for understanding the current state of cosmic acceleration. Notably, for $z \to -1$, $\omega$ equals -1, corresponding to a cosmological constant or a pure DE component.

Therefore, using the expressions above along with Eq. (\ref{w}), we can derive the following results:
\begin{equation}
\label{df}
\frac{3}{\gamma  (1+z)^3+3}-\frac{(2 \beta +3) H^2+(\beta +2) \dot{H}}{(2 \beta +3) H^2+5 \beta  \dot{H}}=0,    
\end{equation}

By employing Eqs. \eqref{zt} and \eqref{df}, we derive the Hubble parameter $H$ as a function of the redshift $z$,
\begin{equation}\label{Hz}
H(z)= H_{0} \left(\frac{-12 \beta +(\beta +2) \gamma  (1+z)^3+6}{-12 \beta+(\beta +2) \gamma +6}\right)^{l},
\end{equation}
where $l= \frac{2 \beta +3}{3 \beta +6}$ and $H_{0}$ represents the current value of the Hubble parameter at $z=0$. When we set $\beta = 0$, the model simplifies to $f(T, \mathcal{T}) = f(T) = T$, establishing a direct connection to the $\Lambda$CDM model. Consequently, the equation for the Hubble parameter $H$ is reduced to
\begin{equation}
H(z)= H_{0} \left(\frac{\gamma  (1+z)^3+3}{\gamma +3}\right)^{\frac{1}{2}},    
\end{equation}
which closely resembles the $\Lambda$CDM model \cite{Mukherjee}.

To describe the accelerated or decelerated nature of cosmic expansion, we introduce the deceleration parameter $q$, defined as
\begin{equation}
\label{q}
q= -1- \frac{\dot{H}}{H^{2}}.
\end{equation}

From Eqs. (\ref{Hz}) and (\ref{q}), we obtain
\begin{equation}
q(z)=-1+\frac{(2 \beta +3) \gamma  (1+z)^3}{-12 \beta +(\beta +2) \gamma  (1+z)^3+6}. 
\end{equation}

The cosmological characteristics of the model are highly sensitive to the values of the parameters $H_0$, $\beta$, and $\gamma$. In the following section, we will use observational data to impose constraints on these model parameters.

\section{Observational constraints and data analysis}\label{sec4}

In this stage, we apply a statistical analysis to compare observable cosmology data with the predictions of the $f(T, \mathcal{T})$ modified gravity. In our investigation, we employ the Pantheon+ sample with 1701 data points and the Cosmic Chronometers dataset with 31 measurements. Also, we estimate the posterior probability using a Bayesian statistical analysis using the likelihood function and the Markov Chain Monte Carlo (MCMC) random sampling method \cite{R26,R27}. For the model parameters, we use flat priors to do the MCMC analysis, choosing physically motivated ranges. In particular, $H_0 \in [50, 100] \, \text{km/s/Mpc}$, $\beta \in [-1.0, 1.0]$, and $\gamma \in [-1.0, 1.0]$ are set. We execute several chains, each with 100000 steps, to guarantee convergence.

\subsection{$H(z)$ dataset}

Cosmic chronometers (CC) are ancient, passively evolving galaxies that have ceased star formation, identified by their distinct spectral and color features \cite{R28}. In this study, we use a dataset of 31 $H(z)$ measurements from these galaxies, covering a redshift range of $0.07 \leq z \leq 2.41$ as provided in \cite{R29}. The $H(z)$ values are derived using the relation $H(z) = -\frac{1}{1+z} \frac{dz}{dt}$, with $\frac{dz}{dt}$ obtained from the observed redshift change $\frac{\Delta z}{\Delta t}$. The CC method’s covariance matrix, $C_{ij}$, combines statistical errors and corrections for younger stellar populations, model dependence, and stellar metallicity uncertainties. Specifically, model uncertainties in $C_{ij}^{model}$ include contributions from the star formation history, initial mass function, stellar library, and synthesis model, structured as follows \cite{COVM}
\begin{equation}\label{4a11}
C_{ij}=  C_{ij}^{SFH} + C_{ij}^{IMF} + C_{ij}^{Ste.Lib} + C_{ij}^{SPS}.
\end{equation}

Furthermore, to conduct the MCMC analysis, it is essential to compute the chi-square function for the CC data, defined as
\begin{equation}
\chi^2_{Hz} = \Delta H C^{-1} \Delta H^T   
\end{equation}
where $\Delta H_i = H_{obs.}(z_i)-H_{model}(z_i)$ and $C$ denotes the covariance matrix.

\subsection{$Pantheon^+$ samples}

The Pantheon+ compilation provides an extensive dataset of 1701 SNe Ia samples across redshifts from 0.001 to 2.3, enabling a detailed study of cosmic expansion. Building on previous compilations like Union \cite{R30}, Union 2 \cite{R31}, Union 2.1 \cite{R32}, JLA \cite{R33}, and $Pantheon$ \cite{R34}, $Pantheon^+$ \cite{R35} includes the latest observations, enhancing its value for cosmological research through precise distance measurements using SNe-Ia as standard candles. The associated $\chi^2$ function is given by 
\begin{equation}\label{4c}
\chi^2_{SNe}=  D^T C^{-1}_{SNe} D.
\end{equation}

Here, $C_{SNe}$ denotes the covariance matrix for the $Pantheon^+$ dataset, capturing both statistical and systematic uncertainties. The vector $D$ is defined as $D = m_{Bi} - M - \mu^{th}(z_i)$, where $m_{Bi}$ is the apparent magnitude and $M$ is the absolute magnitude. In addition, $\mu^{th}(z_i)$ represents the theoretical model's distance modulus, which is calculated using the equation below
\begin{equation}\label{4d}
\mu^{th}(z_i)= 5log_{10} \left[ \frac{D_{L}(z_i)}{1 Mpc}  \right]+25. 
\end{equation}

In this expression, $D_{L}(z)$ represents the luminosity distance, defined as the distance to an object based on its luminosity and the observed brightness of the light it emits. For the specified model. It can be calculated as
\begin{equation}\label{4e}
D_{L}(z)= c(1+z) \int_{0}^{z} \frac{ dx}{H(x,\theta)},
\end{equation}
where $\theta=(H_0, \beta,\gamma)$. In contrast to the $Pantheon$ dataset, the $Pantheon^+$ compilation successfully resolves the degeneracy between $H_0$ and $M$ by reformulating the vector $D$ as
\begin{equation}\label{4f}
\bar{D} = \begin{cases}
     m_{Bi}-M-\mu_i^{Ceph} & i \in \text{Cepheid hosts,} \\
     m_{Bi}-M-\mu^{th}(z_i) & \text{otherwise.}
    \end{cases}   
\end{equation}

In this case, $\mu_i^{Ceph}$ is determined independently using Cepheid calibrators. Therefore, Eq. \eqref{4c} can be rewritten as $\chi^2_{SNe} = \bar{D}^T C^{-1}_{SNe} \bar{D}$. 

Now, our goal is to maximize the total likelihood function $\mathcal{L}_{total}$, derived from the product of individual likelihoods for each dataset: $\mathcal{L}_{total} = \mathcal{L}_{Hz} \times \mathcal{L}_{SNe}$. The total chi-square $\chi^2_{total}$, is minimized and is connected to the likelihood through the relation $\mathcal{L} \propto e^{-\chi^2/2}$. The total chi-square function is calculated by summing the chi-square values of the individual datasets: $\chi^2_{total} = \chi^2_{Hz} + \chi^2_{SNe}$.

\begin{figure}[H]
\centering
\includegraphics[width=9 cm]{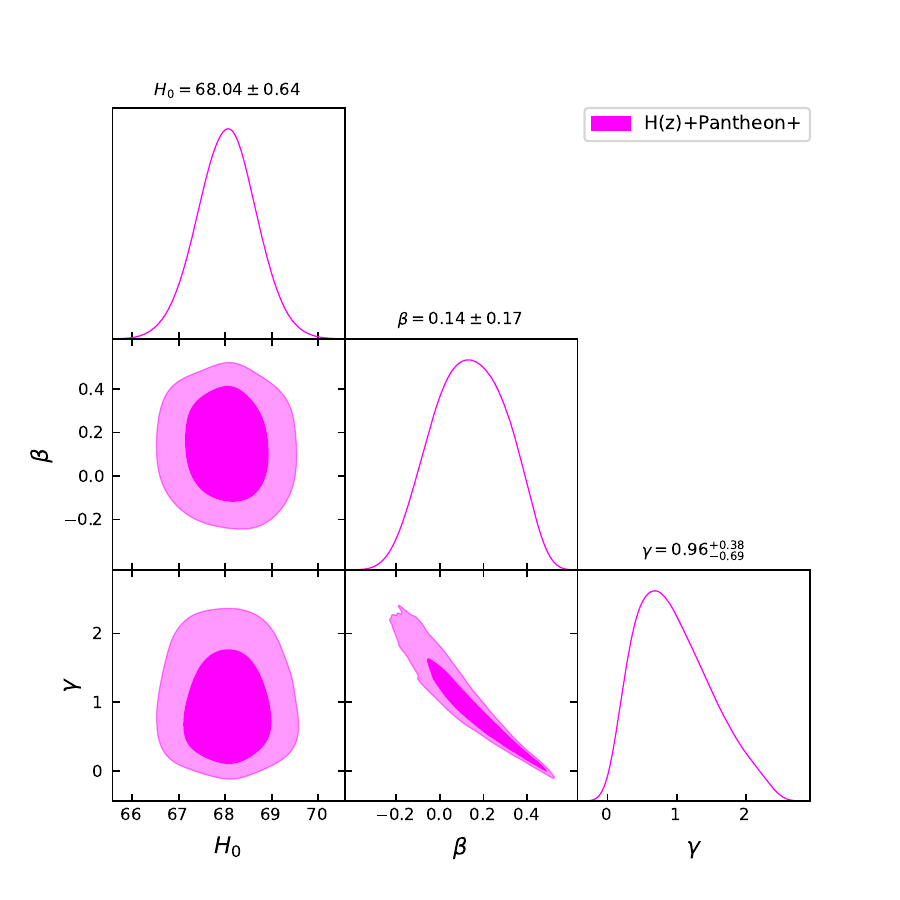}
\caption{The plot shows the best-fit values of the model parameters $H_0$, $ \beta$, and $\gamma$ obtained from the $H(z)+Pantheon^+$ datasets at the $1-\sigma$ and $2-\sigma$ confidence levels.} \label{F_Com}
\end{figure}

In addition, Fig. \ref{F_Com} presents the posterior distributions of these parameters, with the contours representing the 68\% and 95\% confidence intervals. Based on the combined datasets, we can derive meaningful interpretations from the results related to the model. The Hubble parameter is given as $H_0 = 68.04 \pm 0.64$. This value agrees with the latest measurements of the Hubble constant, supporting the current understanding of cosmic expansion and indicating that the universe is expanding at a rate consistent with both local measurements and observations from the CMB \cite{O1,O2,O3}. The parameter $\beta = 0.14 \pm 0.17$ quantifies the strength of the coupling. A positive value of $\beta$ indicates that the gravitational effects are enhanced by the matter content, potentially leading to accelerated expansion scenarios. Furthermore, the value of $\gamma = 0.96^{+0.38}_{-0.69}$ plays a crucial role in determining the rate at which $\omega$ transitions from earlier values (which could be indicative of a matter-dominated universe) to values consistent with DE. A value of $\gamma$ close to 1 suggests a smooth transition, indicating that DE effects became dominant relatively recently in cosmic history.

\section{Discussion of results} \label{sec5}

In this section, we discuss the results obtained, focusing on the behavior of the deceleration parameter, energy density, and the effective EoS parameter using the model parameter values constrained by the $H(z)+Pantheon^+$ analysis.

The deceleration parameter $q$ is a crucial indicator of the universe’s expansion behavior, helping to distinguish between accelerating and decelerating phases. This parameter varies throughout cosmic history, reflecting shifts between different expansion phases. Initially, $q$ was positive in the early universe, indicating deceleration driven by matter dominance. As DE became the dominant component, $q$ shifted to negative values, resulting in the current accelerated expansion. Numerous studies in the literature have utilized the deceleration parameter to explain the evolutionary phases of the universe \cite{DP1, DP2,DP3,DP4,DP5,DP6,DP7,DP8}.

\begin{figure}[H]
\centering
\includegraphics[width=8.5 cm]{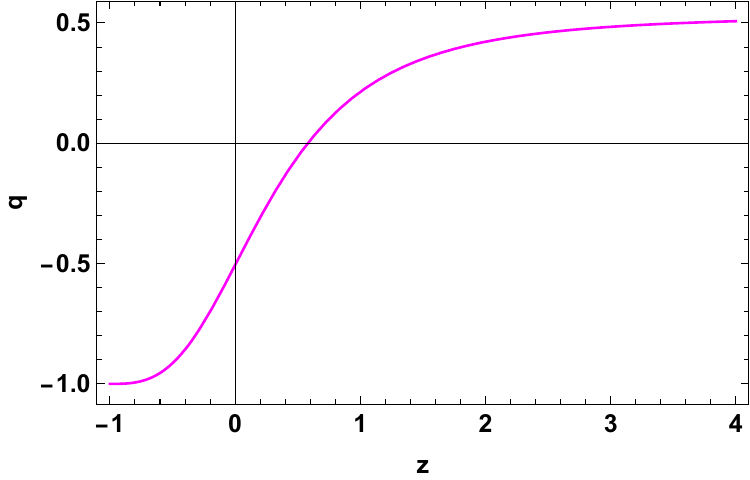}
\caption{The plot shows the variation of the deceleration parameter $q(z)$ using values obtained from the $H(z)+Pantheon^+$ datasets.} \label{F_q}
\end{figure}

Figure \ref{F_q} shows the variation of $q(z)$ based on the constrained values of the model parameters $H_0$, $\beta$, and $\gamma$ obtained from the $H(z)+Pantheon^+$ analysis. In our $f(T, \mathcal{T})$ gravity model, we obtain $q_0 = -0.51$ \cite{q0}, which indicates a universe currently undergoing accelerated expansion. This result is in good agreement with the $\Lambda$CDM prediction, where the deceleration parameter from Planck 2018 results is approximately $q_0 \approx -0.55$ \cite{O3}. The slight deviation between our result and the Planck value highlights the capability of the $f(T, \mathcal{T})$ framework to mimic the late-time cosmic acceleration observed in $\Lambda$CDM while introducing modifications from the extended $f(T, \mathcal{T})$ gravitational model. Such behavior demonstrates the viability of the $f(T, \mathcal{T})$ gravity model as a promising alternative to the $\Lambda$CDM paradigm. The model also yields a transition redshift $z_t = 0.57$ \cite{z1,z2,z3}, marking the switch from deceleration ($q > 0$) to acceleration ($q < 0$), which is supported by observational data. These values help us probe how different values of $q$ through time correspond to different epochs, from the decelerating matter-dominated era to the accelerated dark-energy-dominated era, offering a comprehensive view of the universe's expansion history.

\begin{figure}[H]
\centering
\includegraphics[width=8.5 cm]{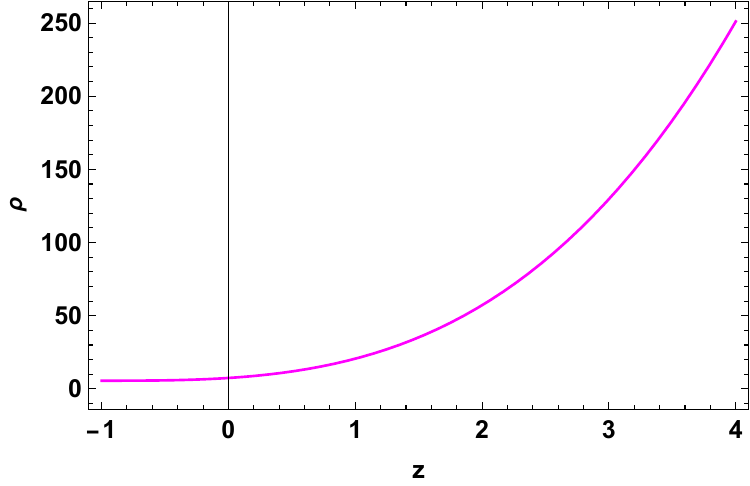}
\caption{The plot shows the variation of the energy density $\rho(z)$ using values obtained from the $H(z)+Pantheon^+$ datasets.} \label{F_rho}
\end{figure}

The EoS parameter plays a crucial role in understanding the dynamics of the universe, particularly in the context of DE and cosmic acceleration. Different values of the EoS parameter correspond to different types of cosmic fluids: for instance, $\omega = 0$ indicates dust matter, $\omega = \frac{1}{3}$ corresponds to radiation, and $\omega = -1$ represents a cosmological constant (vacuum energy). Furthermore, the accelerating phase of the universe, a topic of recent discussion, is indicated when $\omega < -\frac{1}{3}$, encompassing both the quintessence regime ($-1 < \omega < \frac{1}{3}$) and the phantom regime ($\omega < -1$). Moreover, variations in $\omega$ can influence the expansion rate of the universe, the transition from deceleration to acceleration, and the overall fate of cosmic evolution. Understanding the EoS parameter is essential for unraveling the nature of DE and exploring modifications to GR in the realm of cosmology. 

\begin{figure}[H]
\centering
\includegraphics[width=8.5 cm]{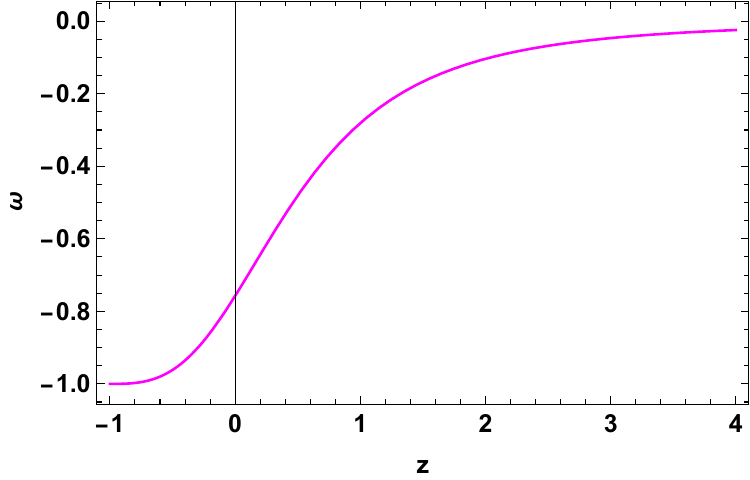}
\caption{The plot shows the variation of the effective EoS parameter $\omega(z)$ using values obtained from the $H(z)+Pantheon^+$ datasets.} \label{F_w}
\end{figure}

Fig. \ref{F_rho} shows that the energy density of the cosmic fluid remains positive throughout the evolution of the universe, indicating a stable, non-negative contribution to cosmic expansion. As time progresses, this density gradually diminishes, approaching zero in the far future, suggesting that the influence of the cosmic fluid on the universe’s expansion will become negligible over time. This trend aligns with models predicting a universe dominated by DE or cosmological constant-like behavior, ultimately leading to a nearly constant rate of expansion. Further, the variation of the EoS parameter in Fig. \ref{F_w} shows that the current value of the EoS parameter (at $z=0$) is $\omega_{0} = -0.76$ \cite{q0}, clearly indicating an accelerating phase consistent with a quintessence-dominated universe.

\section{Conclusion and Future Prospects}\label{sec6}

It is essential to rigorously evaluate emerging theories of gravity to determine their effectiveness in elucidating the dark sector of the universe. One such theory is $f(T,\mathcal{T})$, which offers a compelling framework by incorporating the torsion function $T$ alongside the trace of the energy-momentum tensor $\mathcal{T}$. This innovative approach allows for a deeper understanding of gravitational interactions, particularly in the context of cosmic acceleration. To explore the implications of this theory, we investigated the specific functional form $f(T,\mathcal{T}) = T + \beta \mathcal{T}$, where $\beta$ serves as a free parameter that modulates the influence of matter on the dynamics of spacetime \cite{fTT1,fTT2,fTT3}. This formulation not only opens avenues for examining the coupling between geometry and matter but also enables a comparative analysis with observational data to assess its viability in explaining the universe's expansion and the enigmatic behavior of DE.

Moreover, we employed a well-founded expression for the effective EoS parameter, dependent on redshift $z$, to address the field equations for $H(z)$: $\omega(z) = -\frac{3}{\gamma (z+1)^3 + 3}$, where $\gamma$ is a constant \cite{Ankan/2015}. This formulation effectively represents the late-time accelerated expansion of the universe while also exhibiting characteristics similar to DM in its earlier phases. Using the MCMC method in conjunction with Bayesian analysis, we derived the best-fit values of the model parameters from the joint analysis of the datasets ($H(z)+Pantheon^+$). The key findings include $H_0 = 68.04 \pm 0.64$, $\beta = 0.14 \pm 0.17$, and $\gamma = 0.96^{+0.38}_{-0.69}$. These results indicate a consistent rate of cosmic expansion that aligns with the latest observational data. The analysis of the deceleration parameter, as depicted in Fig. \ref{F_q}, reveals a current value of $q_0 = -0.51$, suggesting a significantly accelerated expansion phase. In addition, the model identifies a transition redshift $z_t = 0.57$, marking the shift from a decelerating to an accelerating universe, corroborated by observational data. Furthermore, Fig. \ref{F_rho} illustrates that the energy density of the cosmic fluid remains positive throughout the universe's evolution, indicating a stable contribution to cosmic expansion. Lastly, the variation of the EoS parameter, shown in Fig. \ref{F_w}, reveals a current value of $\omega_{0} = -0.76$, indicating an accelerating phase consistent with a quintessence-dominated universe. Thus, the findings from this study underscore the potential of the $f(T,\mathcal{T})$ framework as a viable candidate for addressing key questions in cosmology, paving the way for further exploration into the dynamics of the universe and the fundamental principles governing DE.

The present analysis motivates and encourages further study into extensions of the $f(T, \mathcal{T})$ theory, which may represent a viable geometric alternative to DE. Extending our analysis to the case $f(T, \mathcal{T})=\alpha T^n+\beta \mathcal{T}$. This extended form introduces additional degrees of freedom, which could provide a deeper understanding of the dynamics of cosmic acceleration. Specifically, the inclusion of the $T^n$-term allows for deviations from the linear $T$-dependence, leading to potentially richer cosmological behavior. For example, the value of the deceleration parameter $q$ and the effective EoS parameter may exhibit different evolutionary patterns depending on the power $n$. Future work should also incorporate alternative datasets, such as BAO, CMB, and weak lensing, to impose tighter constraints on the model parameters, enhancing the robustness and observational consistency of the results. Moreover, the stability of the $f(T, \mathcal{T})$ theory should be rigorously tested to ensure that it serves as a reliable alternative to $\Lambda$CDM at both the cosmic and perturbative levels. Such investigations could reveal differences in structure formation and cosmic evolution that are distinctive from $\Lambda$CDM, supporting its geometric foundation as an alternative to DE. Future studies could detail stability analyses and compare them at the perturbative level with established cosmological models. Another promising direction lies in the application of $f(T, \mathcal{T})$ to the early-universe inflationary period, assessing the theory’s potential to provide a unified approach that addresses both early and late cosmic acceleration. Furthermore, examining the growth of large-scale structures within this framework would offer a robust test for compatibility with observational data. Finally, extending the theory to non-standard settings, such as anisotropic universes, higher-dimensional cosmologies, or models with additional fields or matter sources, would provide a more comprehensive picture of possible gravitational dynamics under modified gravity. Together, these future investigations will further establish the $f(T, \mathcal{T})$ framework as promising candidates for describing the universe’s dark sector through purely geometric means.

\section*{Acknowledgments}
This research was funded by the Science Committee of the Ministry of Science and Higher Education of the Republic of Kazakhstan (Grant No. AP27510857).


\end{document}